\documentclass[prl,twocolumn,groupedaddress,showpacs]{revtex4}

\usepackage{graphicx}
\usepackage{verbatim}
\usepackage{amsmath,amssymb}

\begin{document}

\title{Ground states and thermal states of the random field Ising model}

\author{Yong Wu}
\author{Jonathan Machta}
\affiliation{Department of Physics, University of Massachusetts,
Amherst, MA 01003}

\date{\today}
\begin{abstract}
The random field Ising model  is studied numerically at both zero
and positive temperature.  Ground states are mapped out in a region
of random and external field strength. Thermal states and
thermodynamic properties are obtained for all temperatures using the
the Wang-Landau algorithm. The specific heat and susceptibility
typically display sharp peaks in the critical region for large
systems and strong disorder. These sharp peaks result from large
domains flipping. For a given realization of disorder, ground states
and thermal states near the critical line are found to be strongly
correlated--a concrete manifestation of the zero temperature fixed
point scenario.
\end{abstract}

\maketitle The random field Ising model (RFIM) is one of the
simplest non-trivial spin models with quenched disorder. Despite
thirty years of study it is still not well understood. It has been
proved that an ordered phase exists for sufficiently low temperature
and dimension $d>2$ \cite{ImMa75,GrMa82,Imbrie84,BrKu87}.  The phase
transition between the ordered and disordered phases for $d>2$ is
believed to be continuous and controlled by a zero temperature fixed
point \cite{BrMo,Villain85,Fish86}. Currently, there is no
controlled renormalization group analysis of the RFIM phase
transition and Monte Carlo simulations
\cite{RiYo,Rieg95,NeBa,MaNeCh00} are restricted to small systems and
have been inconclusive.  As the strength of the random field
increases the transition moves to lower temperature and the critical
line intersects the zero temperature line at a zero temperature
phase transition.  Numerical studies of the zero temperature
transition \cite{Ogielski,HaYo01,MiFi02,DuMa03} play an important
role in understanding the model. Ground states are much easier to
simulate than thermal states and, according to the zero temperature
fixed point hypothesis, the $T=0$ and $T>0$ transitions are in the
same universality class.  Critical exponents have been obtained from
zero temperature studies that are mostly consistent with the
scaling theories~\cite{BrMo,Villain85,Fish86}, series methods~\cite{gof} and real space
renormalization group approaches~\cite{DaScYo,CaMa,FaBeMc}.

In this paper we present numerical results at both $T=0$ and $T>0$
for the same realizations of random fields.  For $T>0$ we use the
Wang-Landau~\cite{WaLa01} and Metropolis algorithms.  For $T=0$ we
find ground states using the push-relabel
algorithm~\cite{Ogielski,GoTa88}.  A major conclusion of the paper
is that spin configurations found near the critical line are
strongly correlated with ground states near the zero temperature
critical point.  This observation is consistent with the original
Imry-Ma analysis, incorporated in the zero temperature fixed point
scenario, that the large scale properties of the critical point
depend on the competition between random fields and couplings with
thermal fluctuations serving only to renormalize the strength of
these couplings.  However,  the correlation found here for single realizations of disorder along the critical line is not implied by the existence of a zero temperature fixed point, which implies only the similarity of zero temperature and positive temperature critical {\em ensembles}.

The Hamiltonian of the RFIM studied in this paper is
\begin{equation} \label{eq:hRFIM}
{\mathcal{H}}=-\sum_{\langle
i,j\rangle}s_{i}s_{j}-\Delta\sum_{i}h_{i}s_{i}-H\sum_{i}s_{i}
\end{equation}
The summation ${\langle i,j\rangle}$ is over all nearest neighbors
$i$ and $j$ on a simple cubic lattice with periodic boundary
conditions, spins $s_i$ take the value $\pm 1$, $\Delta$ is the
strength of disorder, $h_i$ is the random field chosen from a
Gaussian distribution with mean zero and variance one, and $H$ is
the external field. Two important quantities are the magnetization
(order parameter) $m=(1/L^3)\sum_is_i$ and the bond energy
$e=(1/L^3)\sum_{\langle i,j\rangle}s_is_j$. We define the disorder
strength separately from the normalized random fields because one of
our primary concerns is to examine single realizations of random
fields as disorder strength, temperature and external field are
varied.  Previous analytic~\cite{ScSo} and real space renormalization group studies~\cite{DaScYo}  also considered single realizations of disorder at the phase transition but do not compare realizations at different disorder strengths as is done here.

\begin{figure*}
\includegraphics[width=4in]{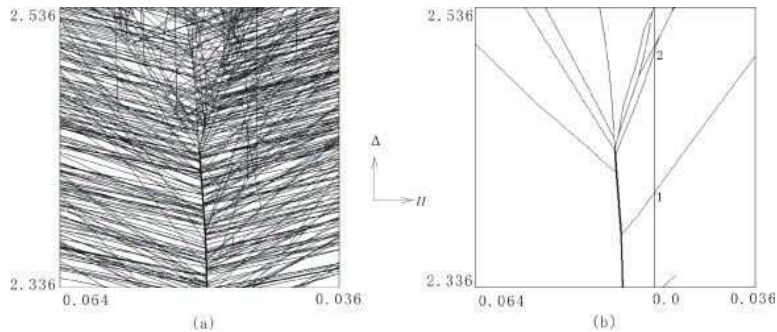}
\caption{Ground states of the RFIM in the $H-\Delta$ plane. (a) All
the ground states of a single $32^3$ realization of disorder. Along
each line two ground states coexist that differ by flipping a single
connected domain. The thickness of a line is proportional to the
magnetization jump across the line.(b) The same realization as in
(a), but only lines with the bond energy jump $\delta e>0.03$ are
shown. Along the $H=0$ axis there are two major jumps, which are
labeled as $1$ and $2$ in the graph.} \label{fig:gndstates}
\end{figure*}

\begin{figure}
\includegraphics*[width=1in]{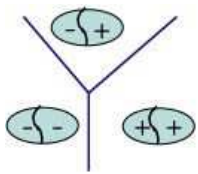} \caption{Schematic picture of a triple point.  The shaded
ovals show the orientation of spins within a single domain that
flips crossing the vertical line and is broken into two pieces
crossing the diagonal lines.} \label{fig:triple}
\end{figure}

Consider the set of ground states of a single realization of
disorder. We obtain these using a method first introduced by
Ogielski \cite{Ogielski}.  To determine the ground state at a given
value of $H$ and $\Delta$, the RFIM problem is mapped onto the
MAXFLOW problem, which is then solved using the push-relabel
algorithm \cite{Ogielski,GoTa88,GoCh97} \footnote{The algorithm is
available from Andrew Goldberg's Network Optimization Library,
http://www.avglab.com/andrew/soft.html}. The set of all ground
states in a region in the $H-\Delta$ plane is mapped out using a
method described in \cite{WuMa} and similar to the techniques
discussed in \cite{DuMa03, FrGoOrVi}. Figure \ref{fig:gndstates}a is
a portrait of all the ground states of a single realization of
random fields in a  $32^3$ system in a small region of the $H-\Delta$
plane near the finite size critical point, discussed below. Each
line represents values of the parameters for which two ground states
are degenerate and across each line a single connected domain is
flipped. Within each polygon bounded  by these lines, a single spin
configuration is the ground state.  At points where two lines cross,
four ground states are degenerate and the four configurations differ
by the orientation of two separate domains.  More interesting are
``triple points'' where a line bifurcates into two lines in a
\textsf{Y} shape. At triple points three ground states are
degenerate but  the three domains corresponding to the three lines
are not independent. The spin configuration at the top of the
\textsf{Y} results from the break-up of the large domain that flips
across the vertical line of the \textsf{Y} as shown schematically in
Fig.\ \ref{fig:triple}. The triple point has some characteristics of
a thermal first-order transition where two ordered states co-exist
with a disordered state.

When a coexistence line is crossed and a domain is flipped physical
quantities except for the total energy are discontinuous. To
visualize the size of the discontinuity, lines are drawn with a
thickness that is proportional to the jump in the magnetization. The
picture is simplified by removing the large number of lines with
small bond energy jump ($\delta e < 0.3$), as shown in Fig.\
\ref{fig:gndstates}b. The simplified picture reveals a tree-like
structure built from triple points.  The triple point with the
largest energy discontinuity is located at the center of the
picture. In the region above this triple point the magnetization is
small while the line extending below the triple point is the
coexistence line separating the plus and minus ordered states.  In
Ref.\ \cite{DuMa03} this triple point was identified as the
finite-size critical point and its scaling properties were studied.
The size of the discontinuity in the bond energy is governed by the
specific heat exponent. We have also examined the large
discontinuities in bond energy and magnetization along the $H=0$ and
shown~\cite{WuMa} that these scale with the specific heat exponent
and magnetic exponents, respectively.  Within a region that shrinks
as $L^{(\alpha + \beta -2/ )\nu}$ and $L^{1/\nu}$ in the $H$ and
$\Delta$ directions, respectively, the tree-like structure is
statistically self-similar but not self-averaging--each realization
has a unique tree--like structure.

\begin{figure*}
\includegraphics[width=4in]{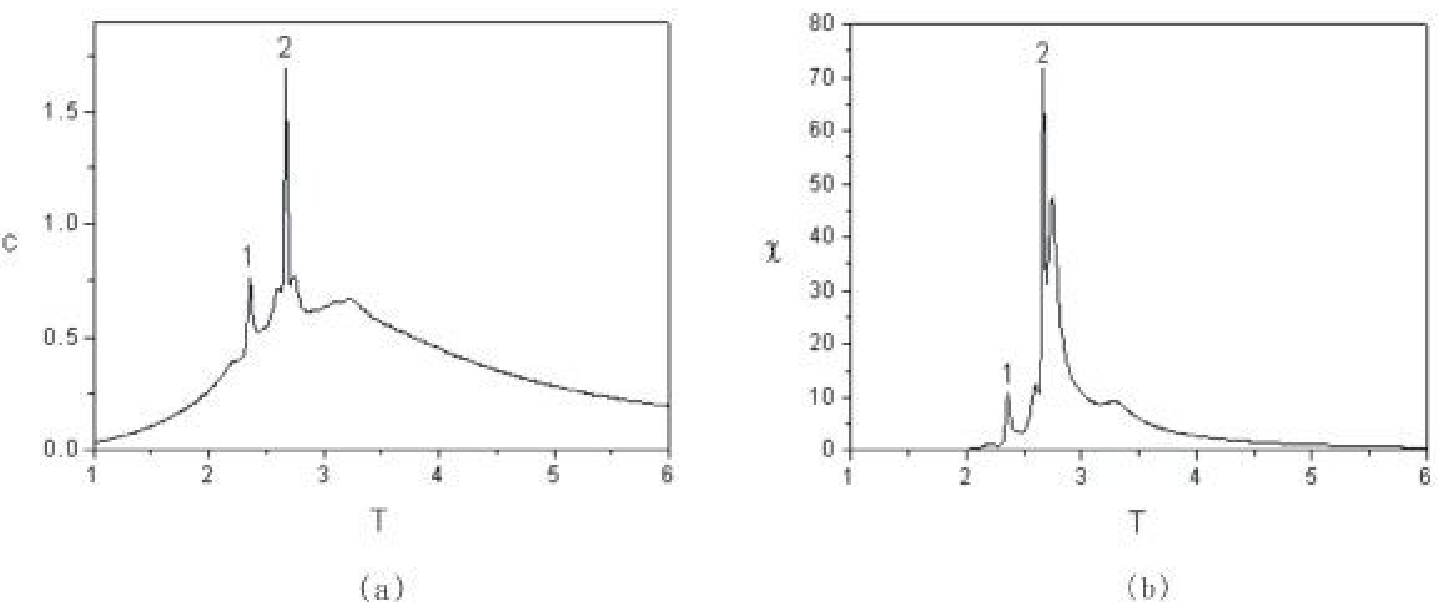}
\caption{The specific heat and the susceptibility of the same $32^3$
realization as in Fig.\ \ref{fig:gndstates} with $\Delta=2.0$ and
$H=0$. Two sharp peaks, labeled $1$ and $2$, are observed, which
correspond to the two large jumps $1$ and $2$ in Fig.\
\ref{fig:gndstates}, respectively.} \label{fig:quant}
\end{figure*}

We study the RFIM as a function of temperature using the Wang-Landau
\cite{WaLa01} and the Metropolis algorithms.  The Wang-Landau
algorithm is a flat histogram Monte Carlo method that automatically
determines the density of states. Thermodynamic quantities at all
temperatures are then derived from the density of states and the
statistics of the magnetization as a function of energy.  The
algorithm smooths the energy landscape and is much more efficient
than the conventional Metropolis algorithm for sweeping a range of
temperatures.  Once a temperature is chosen for detailed study, the
Metropolis algorithm is used to find the thermally averaged spin
configuration. We determined the specific heat and susceptibility
for systems up to size $32^3$.  We find that for large enough
systems ($\geq 16^3$) and strong enough disorder, the specific heat
and the susceptibility typically display one or more sharp peaks. In
Fig.\ \ref{fig:quant} we show the specific heat and the
susceptibility as a function of temperature for the same realization
of normalized random fields whose ground states are shown in Fig.\
\ref{fig:gndstates}.  The random field strength is $\Delta_0=2.0$
and the external field is set to zero. Two sharp peaks appear in
both quantities at the same temperatures. We have simulated one
hundred $16^3$ realizations  with $\Delta_0=1.5$ and find that
about $1/3$ of them have sharp peaks. The number increases to $1/2$
if the random field is strengthened to $\Delta_0=2.0$. For size
$32^3$ and $\Delta_0 = 2.0$ we have simulated nine realization and
sharp peaks are observed for all of them. We tentatively conclude
that the probability of sharp peaks appearing increases with the
system size and the strength of random field.

The sharp peaks in the specific heat and susceptibility can be
understood within the zero temperature fixed point picture of the
RFIM phase transition.  This picture predicts that the behavior in
the critical region at finite temperature is determined by the
competition between couplings and random fields with thermal
fluctuations serving only to renormalize the strength of these
quantities.  One conclusion of this paper is that this scenario
appears to be true for individual realizations of normalized random
fields.  The sharp peaks in the thermodynamic quantities can be
matched one to one with the large jumps at zero temperature.
Furthermore, the spin configurations on either side of the sharp
peaks can be mapped onto the ground states on either side of the
corresponding large jumps.

\begin{figure*}
\includegraphics[width=2.5in]{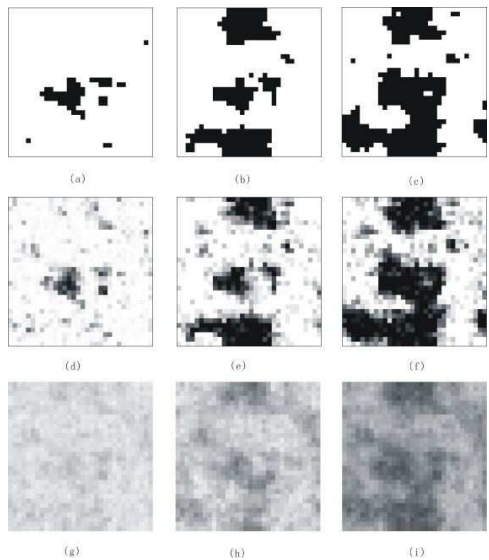}
\caption{Spin configurations near the critical points at zero
temperature and finite temperatures for a single realization of
normalized random fields. Each panel is the same plane of a $32^3$
realization with black representing spin down; white, spin up; and
shades of gray, the thermally averaged spin state.  From left to
right in the top two rows, panels are at $\Delta$ ($T$) before,
between and after jumps (peaks) 1 and 2 in Fig.\ \ref{fig:gndstates}
(Fig.\  \ref{fig:quant}). Specifically, panels (a), (b) and (c) are
ground states at $\Delta=2.36$, $2.41$ and $2.54$, respectively.
Panels (d), (e) and (f) are at $\Delta=2.0$ and $T=2.2$, $2.5$ and
$2.8$, respectively.   Panels (g), (h) and (i) are at $\Delta=0.5$
and temperatures 4.0, 4.3 and 4.45, near the peak in the specific
heat at $T=4.375$. } \label{fig:spin}
\end{figure*}

For a single realization of random fields, we obtain the thermally
averaged spin configuration near the peaks at finite temperature,
and compare these thermal states to the ground states near the two
largest jumps at zero temperature. Figures \ref{fig:spin}d, e and f
show one plane through the system with $\Delta_0 =2.0$ and at
temperatures  just before peak $1$ ($T=2.2$), just after peak $1$
($T=2.5$), and just after peak $2$ ($T=2.8$), respectively. The
difference among the states shows that the sharp peak corresponds to
flipping a relatively large domain.  It is evident that these three
states are strongly correlated with the ground state spin
configuration before the jump $1$ ($\Delta = 2.36$), just after jump
$1$ ($\Delta = 2.41$), and just after jump $2$ ($\Delta = 2.54$), as
shown in Fig.\ \ref{fig:spin}a, b and c, respectively. (The labels
of jumps and peaks are given in Fig.\ \ref{fig:gndstates}b and Fig.\
\ref{fig:quant}.) Similar correlations between ground states and thermal states were found in one dimension \cite{Alava}.

Some correlation between ground states and thermal states persists
to much smaller values of $\Delta_0$ in a regime where the
thermodynamic properties no longer display sharp peaks.  Figure
\ref{fig:spin}g, h and i  show the same realization of disorder and
the same plane through the system but with $\Delta_0=0.5$.  Here the
specific heat has a rounded peak at $T=4.375$. Figures
\ref{fig:spin}g, h and i correspond to temperatures 4.0, 4.3 and
4.45, respectively.  Although there is considerable thermal
``blurring" in these pictures, evidence of the ground state is
unmistakable.

\begin{figure*}
\includegraphics[width=4in]{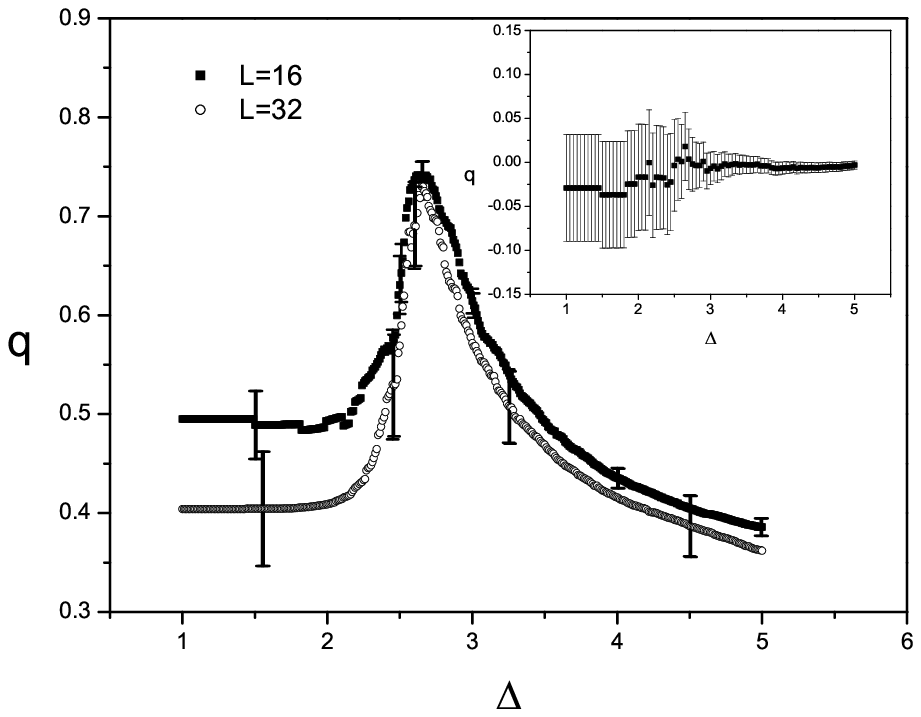}
\caption{Disorder averaged correlation $q$ of a thermal state
just above the transition temperature at $\Delta_0=1.5$ to ground states at disorder stength $\Delta$ for the same realization of random fields.
Solid squares for size $16^3$ and open circles for size $32^3$. Only a
few error bars are drawn to make the figure easy to read. The
inset shows the correlation of thermal states with ground states of a different
random field realization.} \label{fig:corr}
\end{figure*}

A quantitative characterization of the correlation between ground states and thermal states for the same realization can be obtained from the correlation measure,
\begin{equation}
q(\Delta) = \frac{1}{L^3} \sum_i \, \overline{ \operatorname{sgn}(\langle
s_i \rangle_{\Delta,0}\langle
s_i \rangle_{\Delta_0,T^\ast})} 
\end{equation}
where the overbar is an average over realizations of disorder and $\langle s_i \rangle_{\Delta,T}$ is the thermal
average of the spin at the $i$th site at disorder $\Delta$ and temperature $T$ or, if $T=0$, it is the ground state spin value.  For each realization, the temperature $T^\ast=T_{\max} + 0.1$ where $T_{\max}$ is
the temperature of the maximum of the specific heat; one of the sharp peaks in $C$ if sharp peaks exist. Thus, for each realization, we
pick a thermal state just above the transition temperature.
Figure \ref{fig:corr} shows $q$ vs.\ $\Delta$ for
sizes $16^3$ and $32^3$ and $\Delta_0=1.5$, with $96$ realizations for size $16^3$ and $9$ for size $32^3$.  A peak in the correlation occurs at  $\Delta \approx 2.65$ where $q
\approx 0.75$.  The value, $\Delta \approx 2.65$, 
is about $0.15$ larger than the average $\Delta$ at the largest discontinuity in the bond energy for system size $32^3$.  The inset in Fig.\ \ref{fig:corr} shows the average correlation between thermal states of one realization and ground states of another for size $16^3$, which is nearly zero as expected. A second measure, $q^\ast$ is obtained by choosing the value $\Delta^\ast$ for each ground state realization  to give  the largest correlation to the thermal state and then averaging over realizations.  We find that for size $32^3$,  $q^\ast = 0.80 \pm 0.06 $ for $\Delta_0=1.5$ and $q^\ast = 0.85 \pm 0.05$ for $\Delta_0=2.0$.  Together, these result provide quantitative confirmation that the thermal states at temperatures slightly above the thermal critical point are strongly correlated with the ground states at disorder strength slightly higher than the zero temperature critical point. 

The strong correlations between states at different temperatures is
ostensibly in conflict with the idea of ``chaos" in the RFIM. Chaos
in systems with quenched disorder, such as spin glasses and the
RFIM, refers to the sensitivity of spin configurations to small
perturbations either in temperature or in quenched
disorder\cite{BrMo87,AlRi98,MiFi02}.  The existence of chaos in the
RFIM is controversial and is not definitively established.  This
work suggests that chaos is not present along trajectories in the
$\Delta-T$ plane following the critical line.

In summary, we find that sharp peaks in thermodynamic functions resulting from the flipping large
domains are typical near the critical point. In addition, spin configurations near the transition
are similar to the ground states near some corresponding large jump
at zero temperature.  If this connection between critical ground states and
thermal states  persists to large system
size it supports a strong version of the zero temperature fixed
point scenario: the sequence of states near the zero temperature
critical point obtained by varying $\Delta$ for $T=0$ can be mapped
onto the sequence of thermal states near the critical point obtained
by varying $T$ for fixed values of $\Delta_0$, $\Delta_0<\Delta_c$.

\acknowledgements We thank Alan Middleton for useful discussions.
This work was supported by NSF grant DMR-0242402.

\end{document}